# A SURVEY OF EMBEDDED SOFTWARE PROFILING METHODOLOGIES

Rajendra Patel[1] and Arvind Rajawat[2]

[1,2]Department of Electronics and Communication Engineering, Maulana Azad National Institute of Technology, Bhopal, India

rdpatel55@yahoo.co.in[1]

rajawata@manit.ac.in[2]

### ABSTRACT

*Embedded Systems combine one or more processor cores with dedicated logic running on an ASIC or FPGA to meet design goals at reasonable cost. It is achieved by profiling the application with variety of aspects like performance, memory usage, cache hit versus cache miss, energy consumption, etc. Out of these, performance estimation is more important than others. With ever increasing system complexities, it becomes quite necessary to carry out performance estimation of embedded software implemented in a particular processor for fast design space exploration. Such profiled data also guides the designer how to partition the system for Hardware (HW) and Software (SW) environments. In this paper, we propose a classification for currently available Embedded Software Profiling Tools, and we present different academic and industrial approaches in this context. Based on these observations, it will be easy to identify such common principles and needs which are required for a true Software Profiling Tool for a particular application.*

## 1. INTRODUCTION

In the design of embedded systems, design space exploration is performed to satisfy the application requirements. This can be achieved either by different architectural choices or by appropriate task partitioning. At the end the synthesis process generates the final solution with proper combination of Software, Hardware and Communication Structures. Software part may consist of Operating System, Application code and Drivers for peripherals. Similarly the hardware is comprised of appropriate one or more processor cores with dedicated IP cores and communication buses. Design space exploration, from system point of view, can be performed by partitioning the application functionality into hardware and software components. At one step lower level there are multiple processor architectures available for software execution, which are required to be evaluated to identify the most efficient cost effective processor [13]. Processor can also be evaluated for different combination of cache size and bus width and so on [12]. Similarly at the same abstraction level, the hardware components can be explored for different size of FPGAs. Today's System on Chip are mostly realized using higher end FPGAs, which incorporates all three components - hardware, software and communication structures of a system design. In order to achieve a highly cost effective system solution it is very essential to perform performance estimation of Software and Hardware components. In order to increase design space exploration and estimate the software performance, it seems mandatory to use one or the other software profiling tool. Many approaches exists today that claim to provide efficient profiling of the embedded software.

Profiling and Simulation are inherent techniques for acquiring performance information of an application. Simulation offers a great level of accuracy but it is very slow, where as profiling gives fast estimation at cost of accuracy. However, various approaches have been developed which are able to profile application with reasonably good accuracy.

In this paper, we aim to provide an analysis and comparative overview of the state-of-the-art, current directions and future needs in the Software Profiling Tools and Methodologies. We identify common principles based on our observations for classification in Section II. In Section III, we present current profiling approaches researchers have explored in last few years. At last we proposed a general classification and eventually comparison of different profiling tools in Section IV. Finally, the paper concludes with a summary in Section V.

# 2. Classification Of Embedded Software Profiling Techniques

In this section, we will identify common principles in existing embedded software profiling methodologies and develop a classification for such approaches. In this section different criteria based on which profilers can be classified as shown in Fig.1 are discussed briefly. Embedded SW profilers are broadly classified into five categories.

## 2.1 Classification based on Implementation Strategy

According to the strategy of implementation, the profiler can be classified into three further categories as discussed beneath.

### 2.1.1 Software Based Profiling

It is the most common technique for measuring the performance of the application software written using a programming language. There are three different software based profiling methods - insertion of instrumentation code, sampling and cycle accurate simulation.

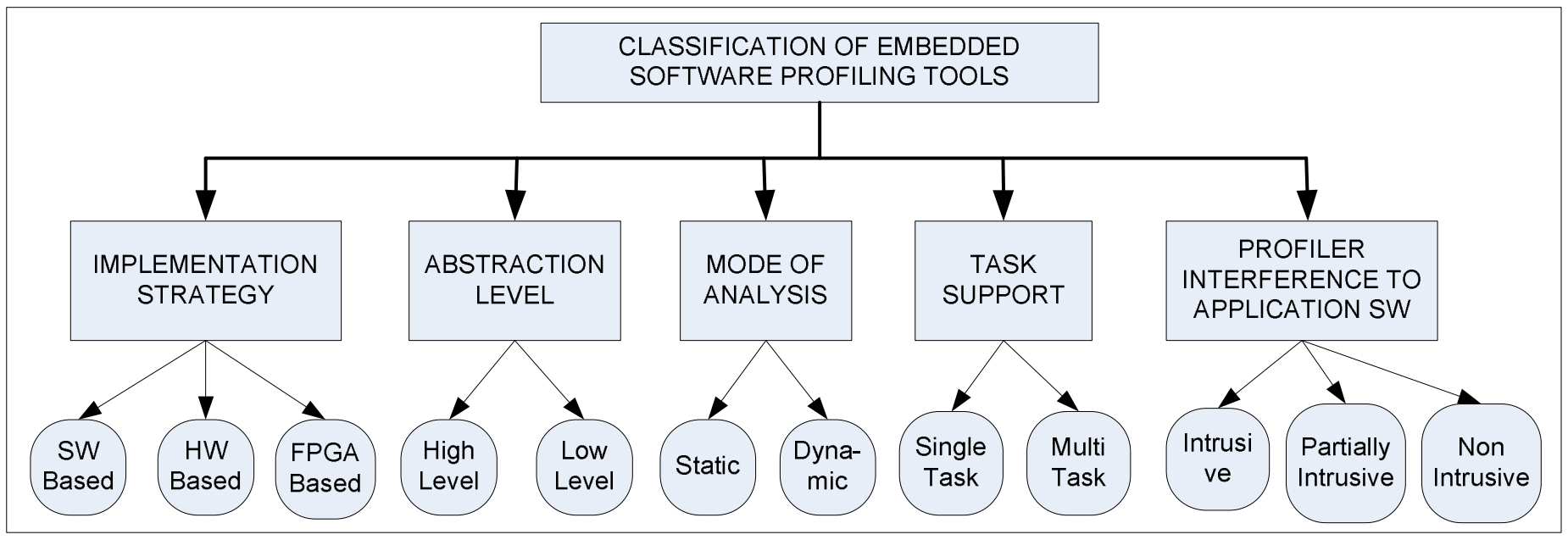

Fig.1 Embedded SW Profiler Classification

*Insertion of Instrumentation Code* – Instrumentation can be done at source code level, assembly code level or binary level [2]. This technique modifies the original code by inserting certain profiling related code. This modified code will be executed on the targeted platform or the host machine. The instrumentation code will help in gathering the profiling information. Execution time for different functions is recorded by software counters running in the profiling tool on the host machine by sampling the PC of target processor at regular interval during program execution. This approach is considerably fast compared to simulation because the executable is running in the real environment. The best example of such profiler is GNU's *gprof*. However the accuracy is significantly poor due to the software overhead introduced by the instrumentation code.

*Sampling* – Statistical sampling [18] is another SW based technique which potentially reduces the runtime overhead compared to previously discussed instrumentation code based SW profiling. In this technique interrupt is generated at a regular interval or a task is written which samples the content of program counter and other important registers of the processor to statically determine execution behaviour latter on.

*Simulation* – For embedded systems, both instrumentation and statistical profiling approaches invariably change the behavior of the application and incur significant runtime overhead. Most of the embedded systems are real-time and designed with tight timing constraints, the minor runtime overhead can lead to missed deadlines and potential system failure. Simulators are nonintrusive and provide accurate profiling information compared to other SW based approaches. The major benefit of simulation is that the designer can track the entire data flow within the internal registers of the processor. The simulation is done on any host machine with the help of Instruction Set Simulation (ISS) model of the target architecture. The accuracy of simulated results depends on how accurate the ISS model is. It is not required to modify the actual executable, hence simulation methods are nonintrusive. Since simulator virtualizes the targeted processor hardware and hence takes from several seconds to minutes to simulate even a few lines of assembly code. Due to such slow behavior, simulation techniques are less attractive for time-to-market critical applications.

### 2.1.2 Hardware Based Profiling

Due to the limitations of software based profiling methodologies, designers have to explore one of the many hardware based profiling techniques. Few of them are Logic Analyzer, JTAG interface and TRACE/DEBUG interface. Logic Analyzer is not suitable for current complex SoCs, which prohibit direct access to a processor's instruction bus. JTAG is useful for validation, verification and debugging; however it is inefficient for profiling an application as it requires significant runtime overhead and changes the execution behavior of the application. As an alternative to JTAG many embedded processors provide TRACE/DEBUG interface that can be utilized to monitor the execution of software application in real time. Traced data is processed with host machine to profile the application code. But to process data in real time requires more powerful host processor compare to processor being profiled.

Hardware counter based profiling (HCBP) [4] is another more promising hardware based profiling technique. This technique makes use of on-chip hardware counters, which are dedicated for profiling purpose. The Performance Advance Programming Interface (PAPI) is required, which facilitates programmers to access these counters at high level interface. PAPI supports range of processors, like Intel Pentium, AMD, Sun Ultrasparc, etc. The hardware counters are dedicated for monitoring events: like memory access, type of instruction being executed, cache miss, pipe stall, etc. HCBP tools do not require the use of instrumentation code and very little performance overhead is introduced during runtime execution. Intel's VTune counter monitor provides an interface for accessing and utilizing the hardware counters to profile application code executing on Pentium based processors.

The *Page Migration Approach* (PMA) [4], developed by Tikir et al. utilizes hardware counters for profiling memory with memory page migrating capabilities. The profiler is used for multi-processor system. Hardware counters are used to sample the frequency at which each processor accesses a page of memory that is remote from the on-board local memory. At a certain numbers of counts specified by the user for remote touching of memory pages, the profiler halts the execution. It moves those particular memory pages to the processor's local memory, for read and writes operations. However, such HW counter based approaches leads to larger area of uP cores and more suitable for higher end processors like Intel's Pentium, AMD, etc.

#### 2.1.3 FPGA Based Profiling

Current embedded systems are widely utilizes FPGAs as implementation platform due to their versatility. Such FPGAs are comprised of hardware customized logic, peripherals along with soft-core processors running on the same chip. FPGA based profiling tools are quite promising compared to the other techniques discussed above for profiling the software application running on such soft-core processors. The profiling tool is implemented on the FPGA near the processor core and collects the profiling information in a nonintrusive manner. There is almost no need of any instrumentation code and very less performance overhead is imparted. Hence such tools provide very accurate profiling outputs.

### 2.2 Classification based on Abstraction Level

Profiling methods can be developed to profile the SW application by collecting profiling data at different abstraction level. In this context profilers can be classified in High-Level versus Low-Level profilers.

*Low-Level Profilers* - Low-Level profiling tools are very close to cycle accurate simulators and estimate the execution time more precisely. In low-level profiler, profiling information is collected during execution of the application code where as in case of high-level profiling; execution of application code is not required. Hence low-level profilers are very slow and do not support fast design space exploration. It is also true that low-level profilers are quite architecture or system dependent and difficult to adapt for other architectures or system.

High-Level Profilers - High-level profiling tools work on formal methods [13][14]. It may comprise of some linear equation relating the count of different type of instructions with total number of clock cycles. However such linear equations do not capture the non-linear behavior of execution due to advance architectural features like pipelines, cache memory and branch predictors. Such nonlinearity can be easily captured using Neural Network techniques. Various Neural Networks can be trained for different domain specific applications and using them performance estimation can be quickly captured at higher abstraction level for fast design space exploration. At the same time high-level profilers are not strongly related to under layered architecture and hence are quite adaptable.

### 2.3 Classification Based on Mode of Operation of Profiler

The profiler can be further classified based on how it collects profiling information. This is another criterion to classify any given profiler, whether execution of the application code is required or not for profiling.

*Static* – In case of static profilers, performance evaluation is done analytically using some mathematical model. Execution of the code is not required.

*Dynamic* – Dynamic profilers analyze the SW application runtime. In case of dynamic profilers,

execution of the application code is essential. Dynamic profilers are good for run time optimization of the system parameters.

### 2.4 Classification based on task supported

Certain profilers are customized to profile single task (thread) applications whereas certain profilers are able to profile multi task (thread) applications [7]. Profilers customized for multitask applications, can also profile Operating System based applications. Multiple threads may be executing on a single processor with OS support or on Multiple Processors [14].

### 2.5 Classification based on profiler's intrusiveness to application code

The profiler may or may not modify the application code in order to perform profiling. Based on this concept, profilers can be classified into three categories.

*Intrusive* – Such profilers greatly modifies the application code and incurs significant execution overhead. Hence accuracy of such profiler's result is significantly poor. However they are fast compared to ISS and are used by people where accuracy is not so important.

*Nonintrusive* – HW and FPGA based profilers are normally nonintrusive. Such profilers pick up the profiling information from program counter and bus activities, without disturbing the processor operation. There is no execution or code size overhead. However, such profilers incur significant area overhead. Such profilers may profile the application in real time and hence offers very good profiling accuracy. They are good for fine tuning the system architecture.

*Partially Intrusive* – Certain profilers slightly modifies the code to record the software execution state. Let say sampling of processor state to profile the software executing on it, is less intrusive than instrumenting the code [4] [18]. However, such profilers give less coverage to profiling than fully intrusive. But they incur less execution and code size overhead.

## 3. Embedded Software Profiling Approaches

Various profiling approaches are explored by different researchers in the world. In [17], Daniel has proposed a profiler to profile large heterogeneous Multi-core Multi-FPGA systems. Such systems are built of multiple boards and each board comprises of multiple parallel software and hardware execution nodes. Hardware resources are in form of FPGAs, whereas PowerPC and Microblaze in form of software processing elements. The main purpose of this profiler is to give insight into the communications occurring between nodes and the computation performed by each node. It is achieved with around 5% performance overhead.

In [13], Oyadama et al. have proposed an integrated approach for software profiling and architecture exploration. Profiling of the software is done using analytical model based on NN Technique. Profiler output helps in selecting the appropriate processor and also guides in Hw/Sw partitioning. Authors demonstrated the approach using ARM processors. The virtual prototype of the entire system, including HW IP, communication bus, processors, is built using MaxSim simulator from ARM. MaxSim provides Bus Functional Model (BFM) of the different components of the system. Hardware components are represented as SystemC modules. At last the performance of the entire system is evaluated by simulating this virtual prototype in MaxSim. Authors have claimed maximum error in performance estimation of SW by 17% with speed up of 35 times with respect to Cycle Accurate simulators. Further improvement is claimed by the same authors in [14].
In this section, we will present various profiling approaches explored so far in academic and industry by different researchers especially in the area of Embedded Software performance estimation. However more famous *gprof,* profiler for general computing software, is discussed as an exception. It is so because number of researchers has compared their results with *gprof.*

## 3.1 GNU's gprof

Profiling using *gprof* [1], [2] can be done using the host system other than the actual target system, which leads to inaccuracies in the profiled data. Hence gprof is more suitable for profiling general computing software than embedded software. As well as profiling speed depends on host system's speed. Changing the host platform affects the Instruction Set Architecture (ISA), the micro architecture and the compiler, which results into variances in the executable that is to be profiled.

To use *gprof*, the application must be compiled with profiling option enabled. Basically, *gprof* looks into each of application functions and inserts code at the head and tail of each one to collect timing information. This code generates an interrupt to sample the Program Counter (PC). Depending on the value of the PC, *gprof* increments the execution time of the corresponding function by the sample time. Thus *gprof* produces statistical results, which implies that until and unless the application execution time is significantly longer than the sampling period, the profiled information will be inaccurate. Hence short executables are run multiple times so that profiling information accumulates for a substantial runtime.

Profiling information is collected by runtime profiler like *gprof,* at cost of overhead in running the profiling software. However, runtime profiling overhead is negligibly small compared to time required to collect cycle accurate information by simulation.

## 3.2 SnoopP Profiler

SnoopP is a real-time, nonintrusive profiling tool which is targeted for soft-core processors instantiated on reconfigurable architectures [3]. It is the FPGA based profiling approach.
*Scope* - It is especially meant for profiling the software application executing on soft-core processors instantiated on the same FPGA, on which SnoopP is implemented. Since SnoopP is nonintrusive, there is no execution overhead. It also provides real-time cycle accurate profiling information faster than cycle accurate simulation.

*Architecture* - The generic SnoopP profiler architecture is presented in Fig.2. The number of code segments depends on the number of functions of the application needs to be profiled. Each segment comprises of two comparators and one counter. Comparators are responsible to check the value of the program counter (PC) between the specified low and high addresses. If the PC value is accessing an address location within the specified bounds, the corresponding counter of that segment will be incremented for each system clock.

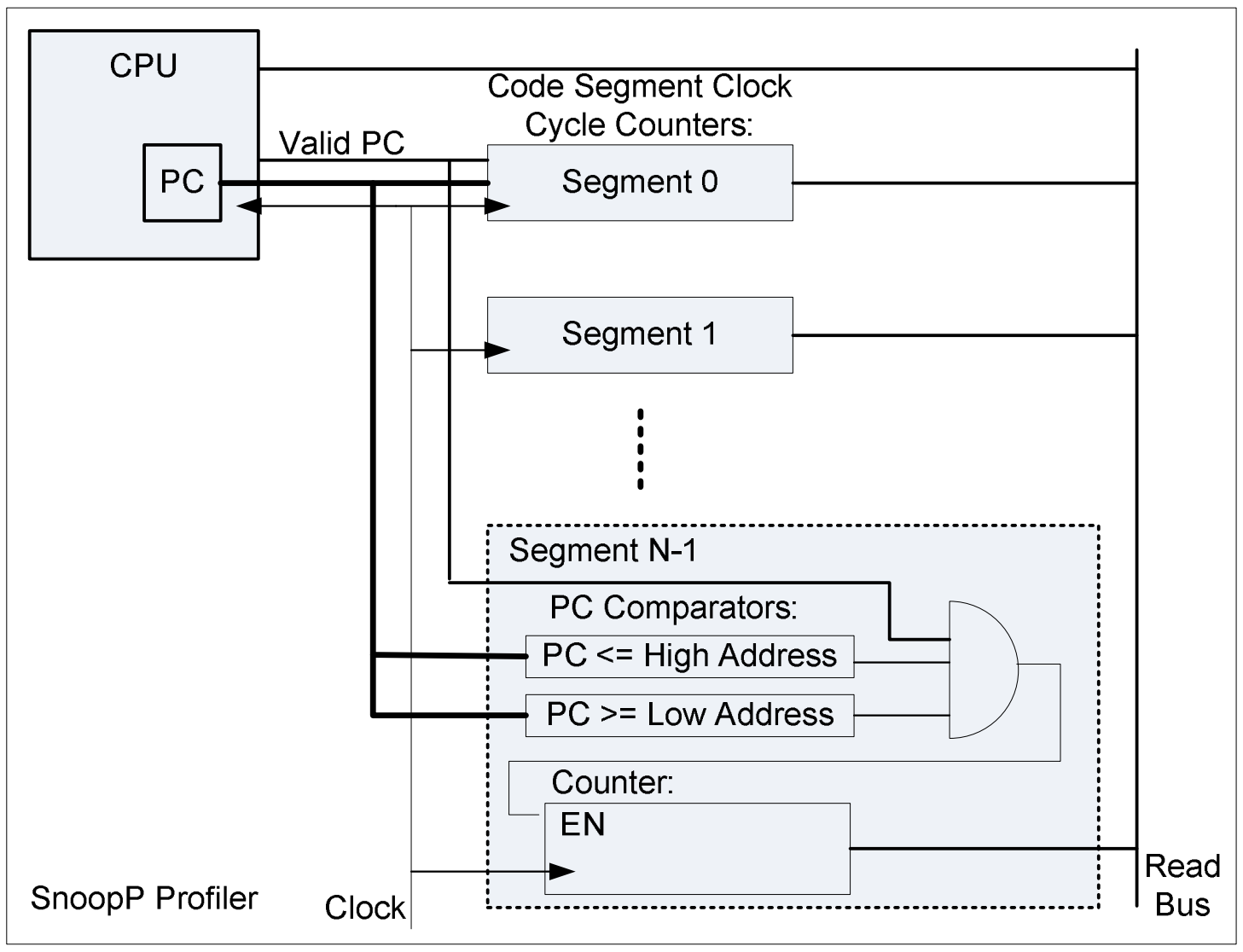

Fig.2 Generic SnoopP Architecture

*Profiling steps* – The entire application is first implemented in software. The functionality is first verified and then the code is compiled to generate the executable. From the assembled code or symbol table the upper and lower bound of addresses for the functions needs to be profiled are extracted. This information is used to determine the number of profiling counters and comparators are set in HDL for each counter. Then the SnoopP is synthesized for the targeted FPGA and profiles the application running on the soft-core processor. If the software implementation fails to meet the design specification, the profiling information is used by the partitioning tool to selectively implement certain components on the hardware. Then the HW/SW co-design is done to comprise hardware and software components. Again, the functional verification is carried out to ensure that the design is correct. The SnoopP is also re-synthesize and used to profile the components running on processor core. This cycle is iterated multiple times until the performance goals are achieved.

*Overhead and Accuracy* - Authors have compare the performance of SnoopP in terms of overhead and accuracy with respect to *gprof*. Dhrystone benchmark is profiled for Microblaze processor on Xilinx FPGA using *gprof* and SnoopP for 100 pass and 1 million pass. SnoopP has produced the results with not more than 0.06% variation, where as gprof has not produce any result for 100 pass. It is so because application has taken less than 10ms on a workstation to execute for 100 pass and this time is very less for *gprof* to collect profiling information. The execution overhead is not more than 0.01% of the total execution time for SnoopP. Fidelity is also observed in the results produce by SnoopP over *gprof* in identifying which functions require the longest execution time.

*Limitation* - The limitation of the SnoopP profiler is that it can profile only the functions which are residing in contiguous locations. If it is required to profile a function X that calls upon sub-functions A,B and C; then four different counters needs to be set to count the number of clocks they consume and letter all are required to add together. Furthermore, if another function Y calls any one of the functions A, B and C; then it will become impossible to differentiate how long it has been executed by function X and Y individually. It is also not capable of counting how many times a function has been called. Another limiting factor is the large area overhead of SnoopP due to large counters to hold profiling information.

### 3.3 Airwolf

Airwolf is also on-chip, real time, FPGA based software profiler, developed for Nios II processor to be synthesized on Altera FPGAs [4]. Contrary to gprof, Airwolf does not require to modify the executable of the application code. But Airwolf inserts extra code to each function before compilation, which is also known as source code instrumentation. A pair of software drivers is added around the software function block. Role of these drivers is to enable or disable a particular profiling counter in the Airwolf. This minimally disturbs the program and software behaviour during execution. Hence, Airwolf seems to be partially intrusive in nature.

*Scope* – The scope of this approach is limited to profiling performance of a software application running on a single soft-core processor on Altera FPGA. However, the methodology can be easily adopted for other processor and FPGA. Airwolf cannot profile memory related events.

*Architecture* – Airwolf contains 20 profiling counters as shown in Fig.3, which support profiling of up to 20 functions at a time. Each profiling counter consists of two counters: one is 32-bit hit counter and the other is 64-bit time counter. Time counter keeps track of execution time of the function and hit counter records how many times the function has been called. Hit Counter counts the positive edges on HCEN control signal and Time Counter counts the number of system clocks for the function being profiled when it is enabled by TCEN signal.

*Overhead and Accuracy* – Authors have claim 66.2% improvement in accuracy in profiled results and 41.3% reduction in runtime performance overhead compared to software based gprof profiler. Authors of Address Tracer [5] have also compared their design with Airwolf and reported a marginal overhead of 0.23% in Airwolf over Address Tracer.

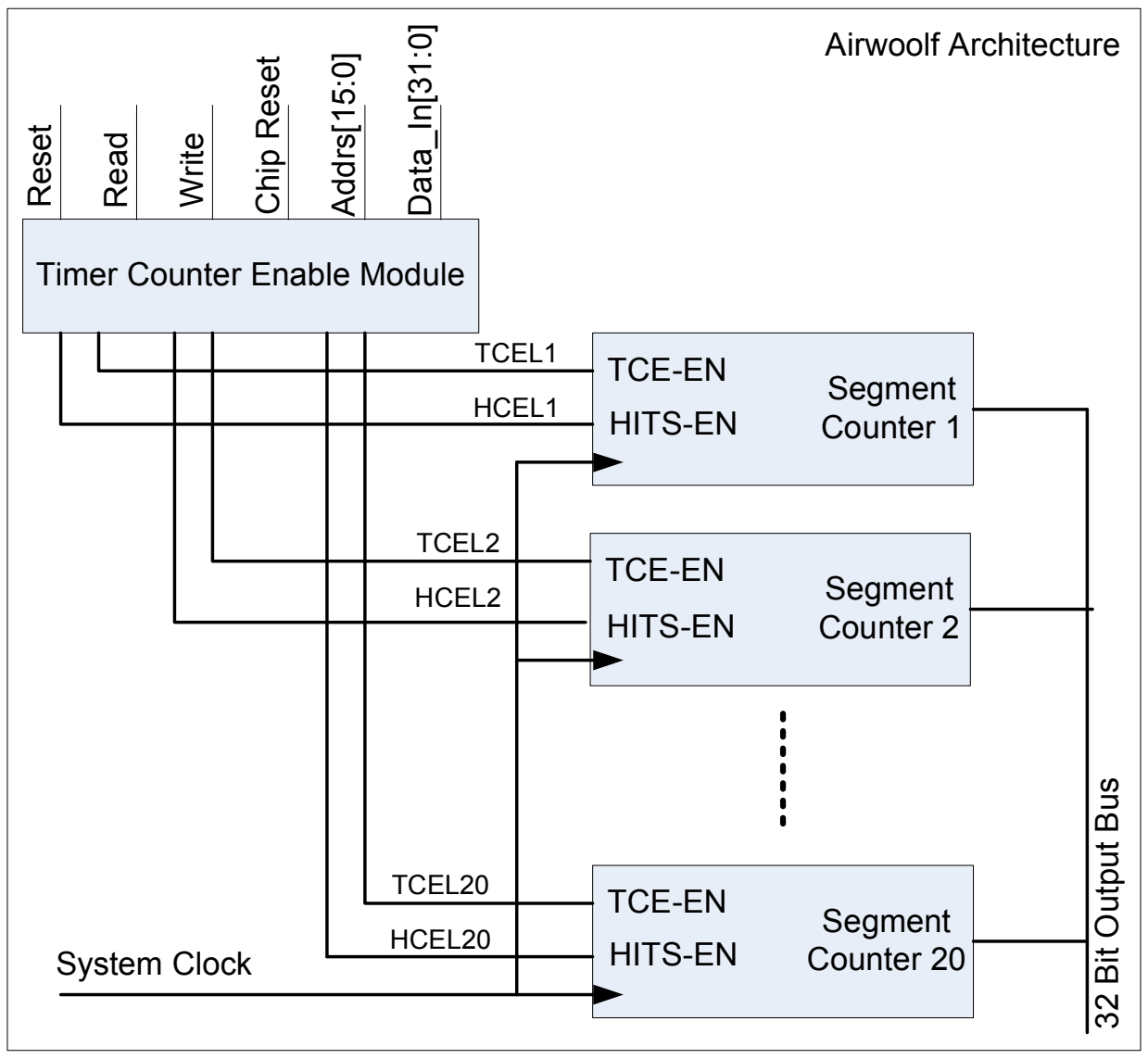

Fig.3 Airwolf Architecture

*Limitation* – The profiled data generated by Airwolf will be erroneous due to the software driver which needs to be executed before the function returns. Hence the clocks consumed in returning the functions are ignored. Similarly driver at the starting of the function is inserted after the header of the function. Hence the system clocks consumed in initialization of header will be ignored by the profiler. This leads to error in counting number of clock cycles for the corresponding function. This error will be significant for short functions.

## 3.4 Address Tracer

Address Tracer [5] is the combination of the SnoopP and Airwolf. It is fully non-intrusive FPGA based profiler. Apart from profiling the given function as discussed in SnoopP, Address Tracer can also keeps track on how many times a function is called. It can also profile the functions correctly even if two different functions call the same function, which is not possible with SnoopP.

*Scope* – The scope of the methodology is found for profiling the software running on soft-core processors in FPGA.

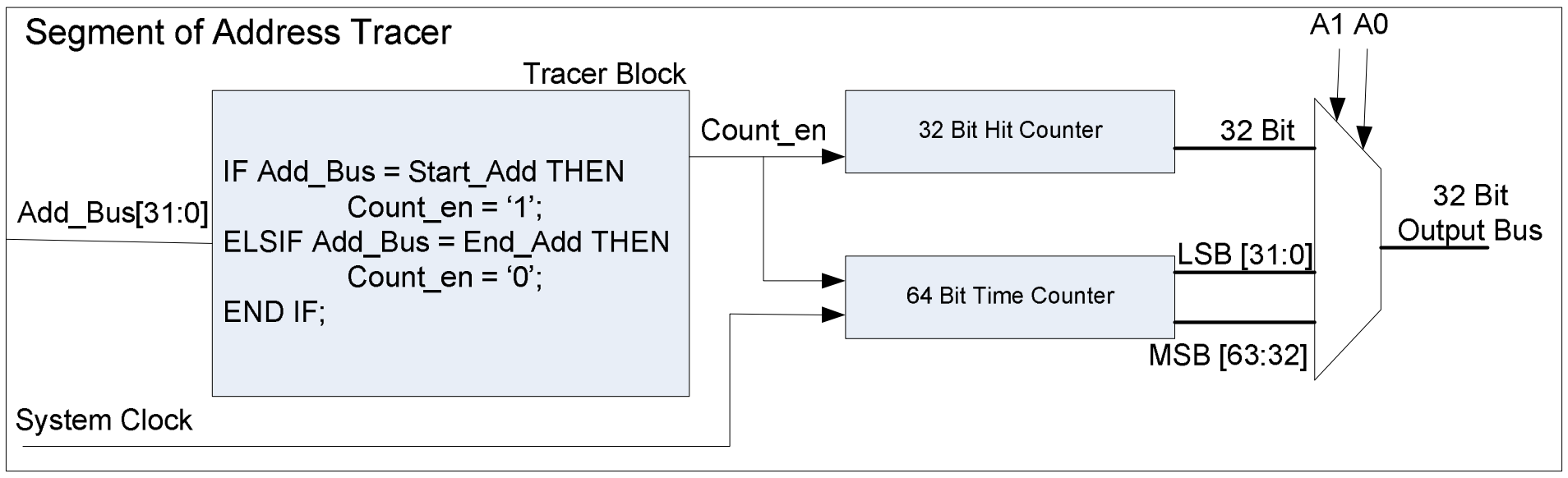

Fig.4 Architecture of a Segment in Address Tracer

*Architecture* – Each piece of code required to be profiled is referenced as segment and a pair of counters is allotted to each such segment to facilitate profiling. The segment counter in Address Tracer is comprised of two counters: 32 bit hit counter and 64 bit time counter as in Airwolf. General architecture of each segment is shown in Fig.4.

*Overhead and Accuracy* – There is no significant change in profiled results of Address Tracer over Airwolf. However the code size of each function is larger in case of Airwolf due to software drivers inserted in each function compared to Address Tracer. Hence a marginal 0.23% performance overhead is observed in Airwolf compared to Address Tracer and in case of gprof it is 20.37%. There is no performance overhead in case of Address Tracer.

*Limitation* – The area of the profiler increases with increased number of functions to be profiled. Another limitation is that it cannot profile memory accesses. Profiling of multiprocessor system is also not supported.

## 3.5 FLAT

Frequent loop analysis tool developed by Ann and Vahid [6], is the first ever dynamic software optimization tool meant for embedded software performance analysis. It is very size and power efficient and fully nonintrusive to software execution. This tool not only detects highly critical regions of the software but also provides the relative frequencies of execution for different loops. This feature is highly important for HW/SW Partitioning.

Motivation behind development of FLAT was an observation that 51% of execution time was occupied by loops having less than 32 instructions in one of the standard benchmark. Hence the problem of critical region detection is actually the problem of detecting frequently executing loops and subroutines for software optimization and HW/SW partitioning.

*Scope* – The scope of the FLAT approach is found to profile any SW application in SoC platform on a FPGA. It is suitable for any embedded system application.

*Architecture* - It is cached-based architecture as shown in Fig.5. The architecture of frequent loop detector employs a 2-way set-associative 32-entry cache, with each entry capable of storing 32 bit frequency counter. The cache is used to store frequency counts of different critical loops and is indexed into using sbb instruction addresses. Here sbb is representing any short backward jump instruction. Whenever any loop counter saturates, the cache contents for all counters will be right shifted by one bit position and hence maintain the relative frequency. The frequent loop cache controller handles the operation of the frequent loop cache. When the sbb signal is asserted, a read of the frequent loop cache is done using the sbb address as the index. If the result is a hit, the frequency is read from the cache, incremented and written back in the next cycle. If the result is a compulsory miss, the instruction is added to the cache with a frequency data value of one. If there is a conflict miss, the new address replaces the old address in the cache with a frequency of one. Set Associative cache is implemented to allow for multiple frequent loops to be mapped to the same set without conflict. If conflicts still occur, the replacement policy will replace the least frequent value in the set with the new incoming sbb.

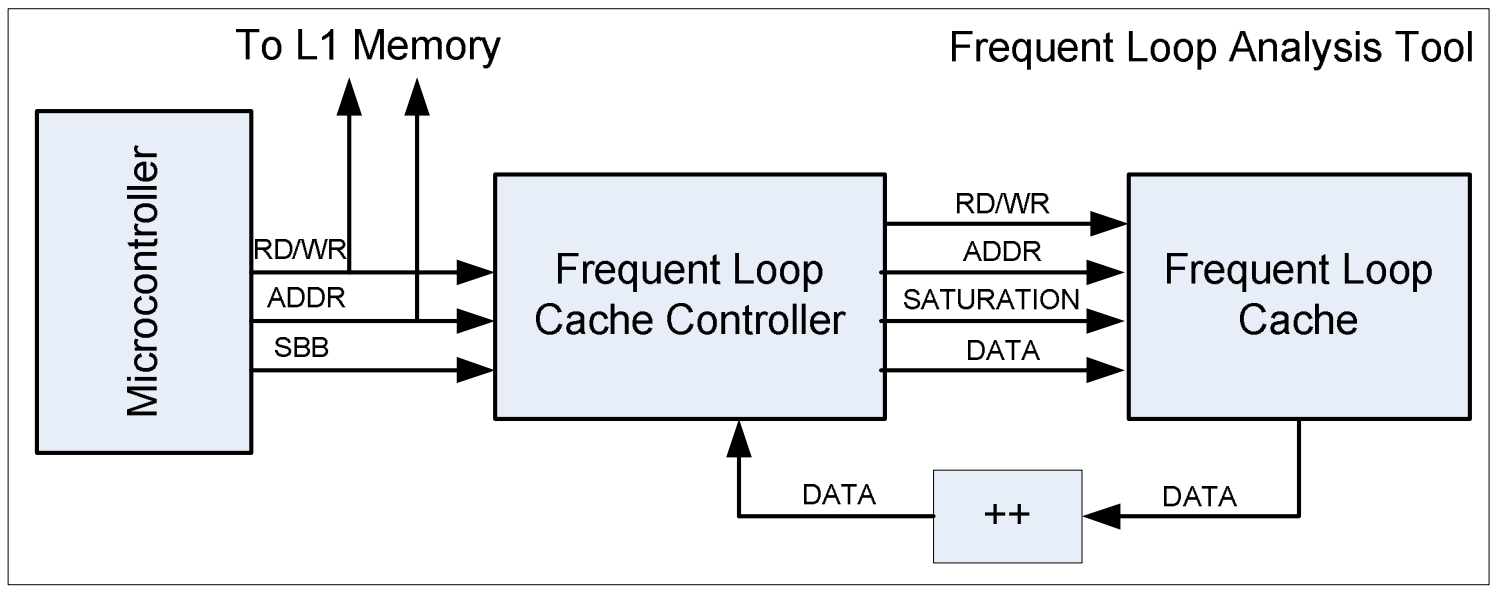

Fig.5 Frequent Loop Analysis Tool Architecture

*Overhead and Accuracy* - Authors have represented a small area, power efficient on-chip profiling architecture to detect the most frequent loops of the software application in a nonintrusive manner. The proposed detector's versatility is verified and validated by 19 embedded system benchmarks. The power overhead is hardly 1% to 2% and can be minimized to even 0.02% compared to 32 bit MIPS 4Kp processor using coalescing and sampling methods. The resulting area overhead is 6.68 % to 12.8 % compared to reported area of MIPS 4Kp processor. There is no runtime overhead. The accuracy depends on the size of the on-chip cache in the FPGA and it is at least 80% for most of the benchmarks used except two for which accuracy is 72%.

*Limitation* – FLAT is developed on the basis that all inner loops employs short backward branch (sbb) and the tool detects such sbb and profiles the application to identify most costlier loops. However, unstructured assembly code generated by hand or by compiler optimization, could result in loops with different structures. Such loops cannot be profiled by the FLAT tool. One more point to be noticed is that all soft-core processor needs to be modified to generate *sbb* signal, whenever a short backward branch instruction is executed.

### 3.6 DAProf

DAProf stands for Dynamic Application Profiler. The first architecture of DAProf proposed in [9] is targeted to characterize the most costly 32 loops in the application, which is more than enough for current average complex application. This design was supporting only single threaded application. In [10] DAProf is extended for multitask support, too. This version of DAProf is capable of profiling an executing application by monitoring the application's short backward branches, function calls, function returns, as well as context switches. Using this information it is possible to characterize frequently executed loops within multitasked applications. It implies that DAProf is capable of profiling RTOS based embedded software. However in latest publication [11], authors have produced more regress results and highlight that DAProf is not up to the mark to profile multitask applications. Interface of DAProf with microprocessor is shown in Fig.6.
*Scope* – DAProf is nonintrusive, low level, dynamic, SoC (ASIC) based profiler which is customized to profile ARM processor based application. Profiler Cache is a small memory, which maintains the current profiling results and intermediate information needed for loop identification, iteration and execution profiling statistics. Profiler Controller is ultimately responsible for determining which profile cache entry should be replaced when new loops are executed.

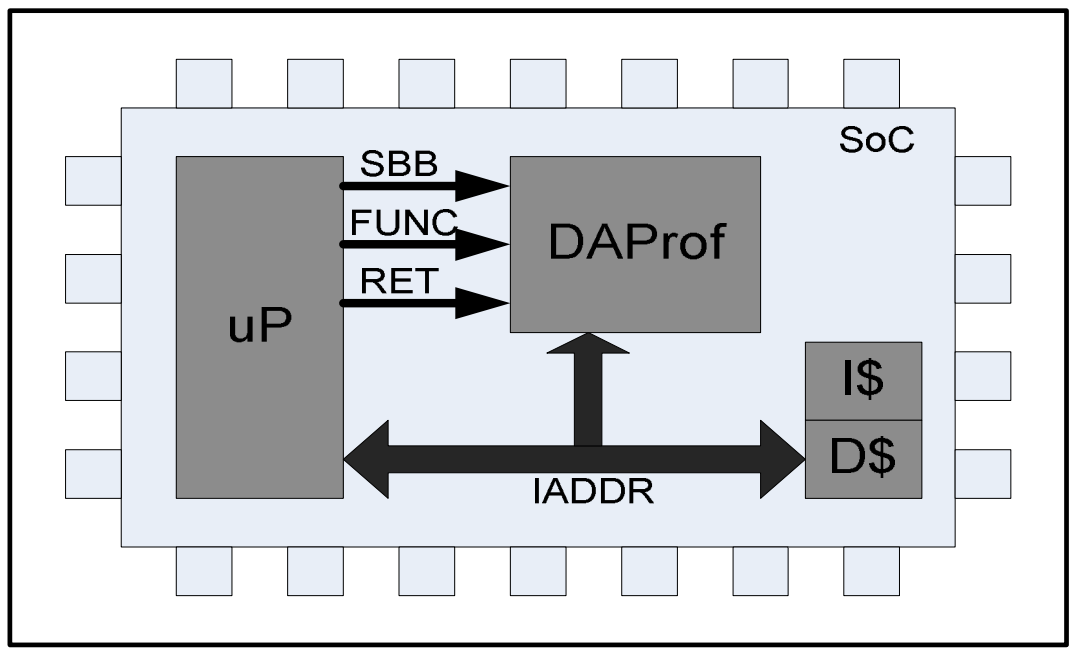

Fig.6 DAProf Interface with Microprocessor

*Architecture* – DAProf is mainly consisting of three functional units as shown in Fig.7. First is Profiler Task Filter and second is Profiler FIFO and the third is Profiler Controller. The profiler task filter provides great flexibility in profiling a multitasked application by allowing designer the option to selectively profile specific tasks, functions, library code, system calls, etc., while filtering out those elements which are not of immediate interest. The profiler FIFO stores the address of interest, short backward branch offset and an encoding indicating the nature of profiling event. In addition Profiler FIFO is also responsible for synchronizing between the operating frequency of microprocessor and Profiler Task Filter and internal design of DAProf.

*Overhead and Accuracy* – Authors have presented an exhaustive analysis of area overhead versus accuracy of DAProf profiler. DAProf with and without function call support have been analyzed. Authors have claimed 98% accuracy in loop execution count and 95% accuracy in estimated execution time. This accuracy is reported in DAProf with function call support and fully associative cache to profile the tasks in a software application with as little as 20% area overhead, on SoC synthesized using UMC 0.18 Micron Library, compared to an area covered by an ARM9 processor executing at 533MHz. Authors have also reported comparative performance of DAProf which covers 76% of the total execution time over FLAT tool that covers only 62%.

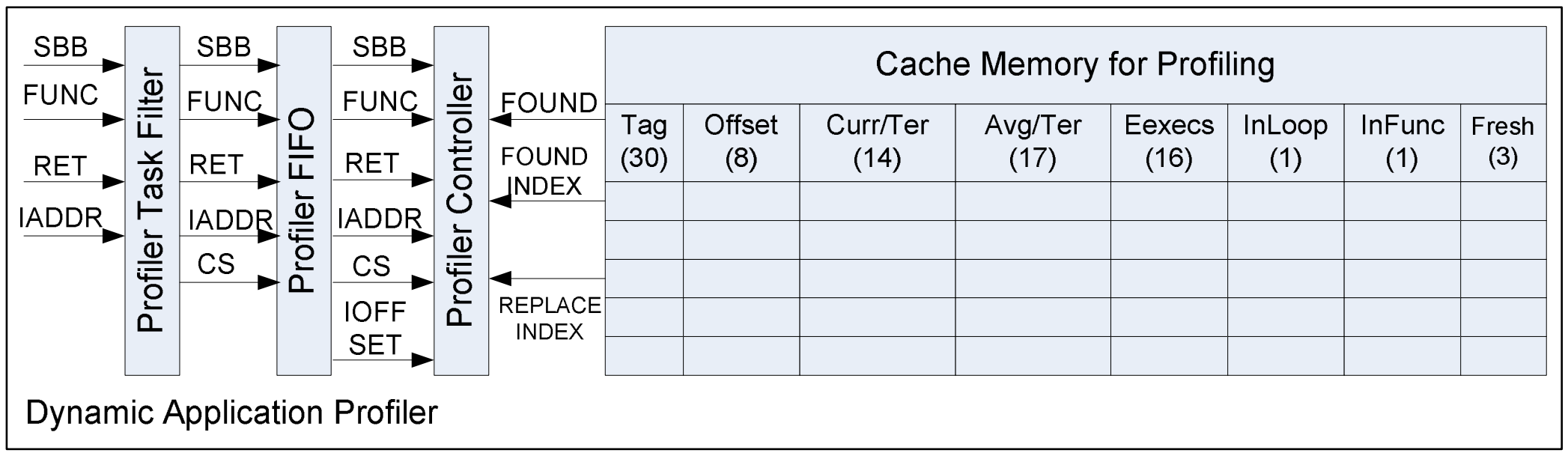

Fig.7 Functional Units of DAProf

*Limitation* – It is required to modify the original processor core to provide profiling event signals to DAProf, like SBB, FUNC and RET. Another limitation of DAProf is that it is more suitable for single threaded application. In case of multitasking applications context switches can leads to reduced accuracy of the DAProf and hence more efforts are required to investigate other possibilities to extend DAProf to profile multitasking applications.

## 3.7 MPPA

MPPA stands for **M**ulti**p**rocessor **P**rofiling **A**rchitecture and is proposed by Po-Hui Chen et al. for MPSoC embedded systems [7].

*Scope* – This approach is suitable for monitoring all processors and system wide events in MPSoC environment on an FPGA. The integration of MPPA and LION3 processors is shown in Fig.8. Since there are no architecture dependent steps, this methodology can be followed for any other profiling framework for any architecture.

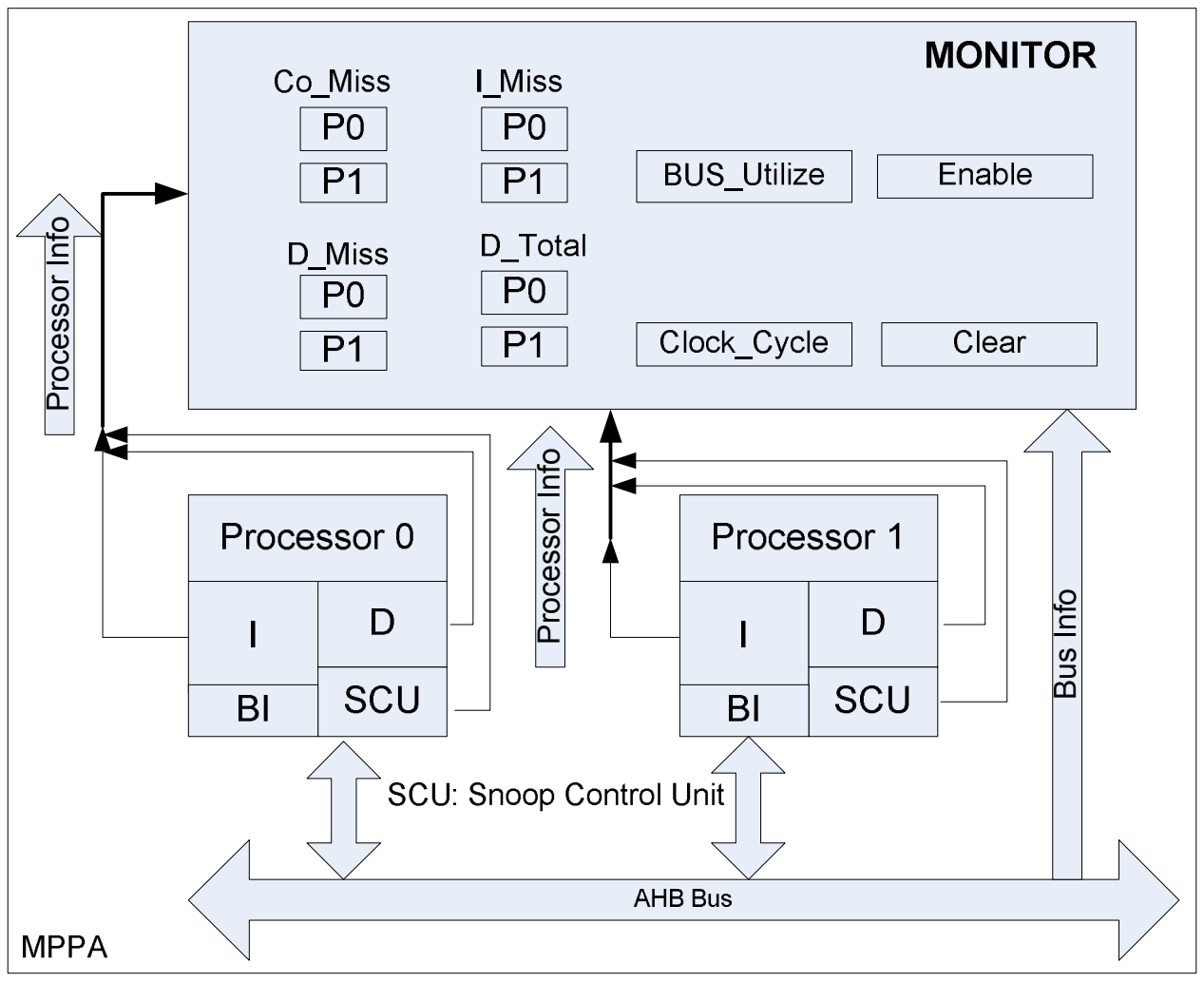

Fig.8 Integration of MPPA and LION3 based system

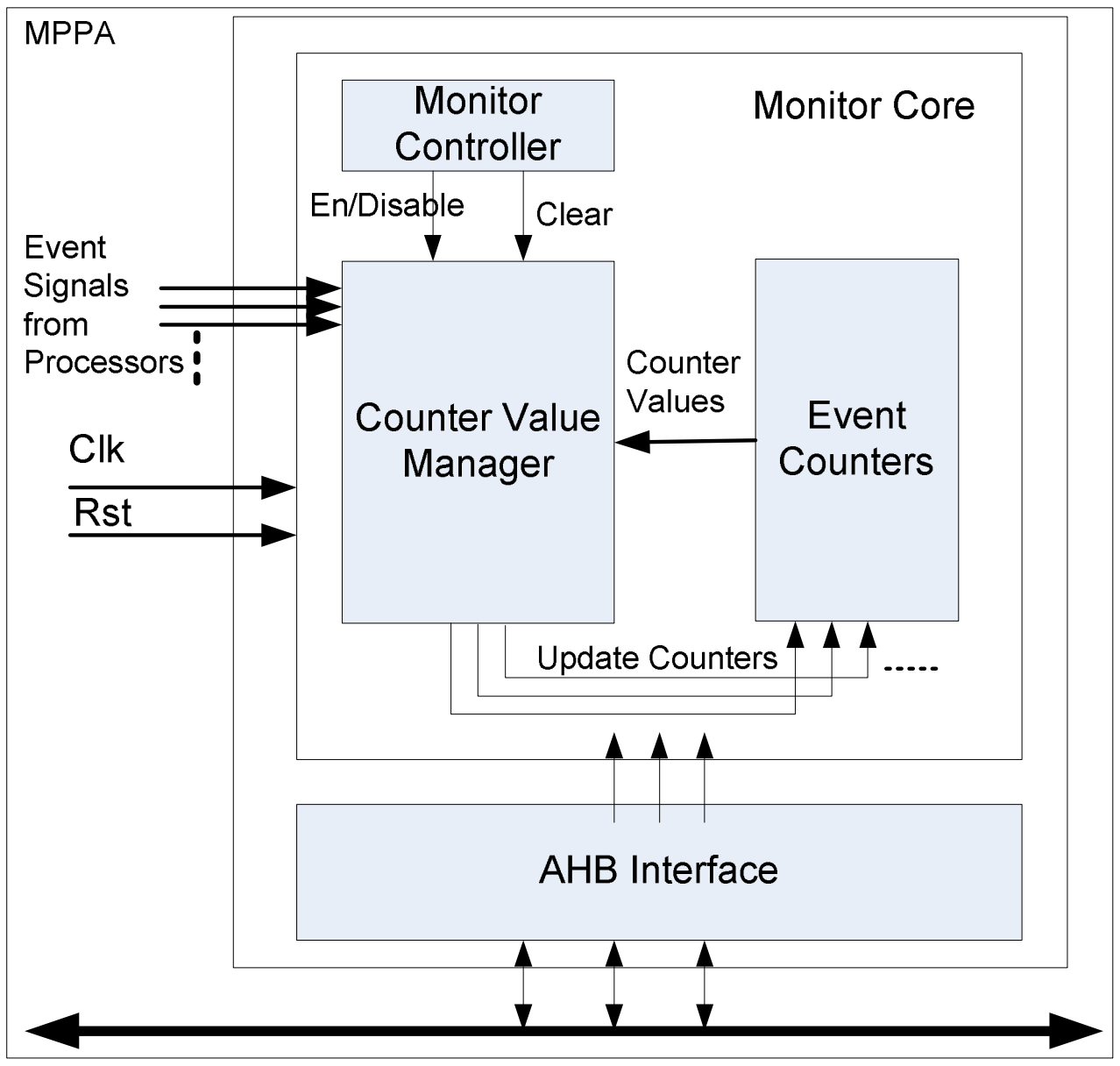

Fig.9 MPPA Profiler Block Diagram

*Architecture* – MPPA is consisted of two parts: Event Sensing and Event Collecting, as shown in Fig.9. Event sensors are embedded in the components and interconnect to sense low level events. All event sensors pass the occurrence of profiling event to the monitor unit where it will be recorded in the profiling counters. Proposed MPPA profiler is implemented for LION3 based MPSoC system on a single Virtex5 FPGA. LION3 is, open source, synthesizable soft processor core based on SPARC V8 architecture with 7 stage pipeline and multiprocessor support. AMBA-2.0 AHB/APB bus interface is used as communication bus.

Linux is used as OS for the system and device driver is written for the MPPA hardware. Monitor module of the profiler is made of two major blocks: Bus Interface and Monitor Core, as shown in Fig.9. All performance counters are memory mapped. Hence any processor can read these counters and make it available to the user for analysis. And no extra interface is needed to be developed. These counters are accessed in Linux OS by writing device drivers instead of system calls that improve cache performance and scalability.

*Overhead and Accuracy* – Authors have claimed only 0.66% gate count overhead by the profiler hardware. Performance statistics, in terms of CPU cycles, for a dual threaded program which processes a large integer array have been estimated by the proposed profiler. The results are found 93.41% accurate against manually calculated cycles.

*Limitation* – Profiling counter overflow is not handled in the present MPPA approach. This may leads to inconsistent results. In large multiprocessor systems often it is required to gather profile information upon occurrence of certain event. However, there is no interrupt (event) or time based sampling of profiling data.

## 3.8 DPOP Framework

DPOP stands for Dynamic Profiling and Optimization. Dynamic profiling opens opportunity to monitor how system responds to changes in environmental conditions or changes in the underlying platform. Effectiveness of the methodology is demonstrated in reference to Forest Fire Detection and Propagation Tracking and Building Monitor applications [8]. Authors have proposed four different dynamic profiling methods suitable for applications with diverse behavior.

*Scope* – Dynamic profiling and optimization is required in many applications like HW/SW partitioning, Energy optimization, Disaster response application, etc. DPOP framework is customized for performance optimization of sensor nodes of wireless sensor network (WSN).

*Architecture* – For dynamically profiling a sensor-based application, DPOP environment requires profiling methods to be incorporated within each node to monitor the execution behavior for individual sensor nodes. At the same time, to optimize the entire sensor-based system, a global view of the entire system is also needed. However, in current DPOP framework profiling at individual node is supported. The broad picture of the framework is as shown in Fig.10. It is comprised of Profiler and Optimizer. Profiler collects the vital runtime performance statistics. Optimizer evaluates these statistics by comparing with end user design metric specification and tunes the behavior and architecture of underline node in order to minimize the overall cost.

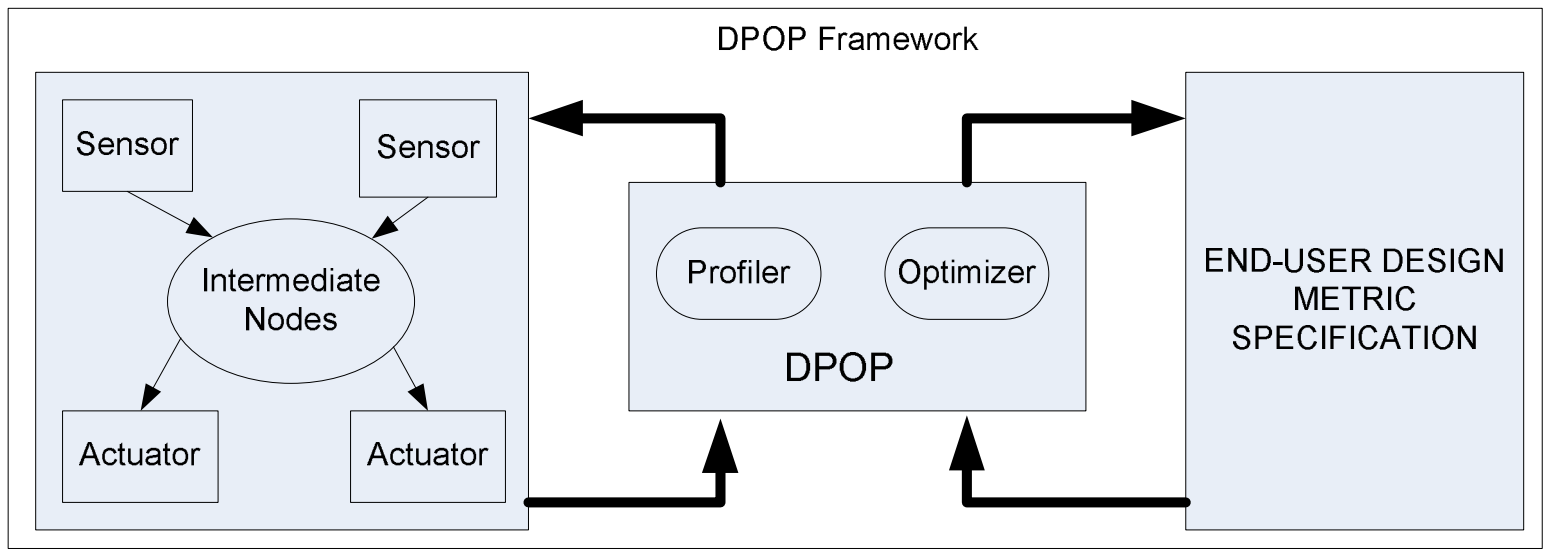

Fig.10 DPOP Framework

*Overhead and Accuracy* – Profiler is aimed to optimize each node independently, a cluster of nodes or entire network in order to minimize energy consumption without incurring significant code size and network traffic overhead. Authors have proposed four different profilers with diverse strategies as depicted in Table 1. In case of Fire Detection and Propagation Tracking application profiling method PM3 yields the lowest traffic overhead of 7.9%. And in case of Building Monitor application, PM1 incurs lowest traffic overhead of 11.6%. Across all profiling methodologies, energy and code size overhead remains reasonably low with maximum of 0.06 mA-Hr and 3.5% respectively. On an average, DPOP can yield up to 83% improvement in overall design cost compared to statically optimized node configuration.

TABLE I
PROFILING STRATEGIES IN DPOP ENVIRONMENT

| **Profiling Strategy** | | **PM1** | **PM2** | **PM3** | **PM4** |
|---|---|---|---|---|---|
| What to profile | Sensor sampling rate | ● | ● | ● | ● |
| | Time between successive packets | ● | ● | ● | |
| Whom to profile | Individual Nodes | ● | ● | ● | ● |
| When to profile? | Profiler module directed | ● | ● | | ● |
| | Periodic | | | ● | |
| How to profile? | Piggybacked | ● | | | |
| | Separate profile packets | | ● | ● | ● |

*Limitation* – The proposed DPOP framework is limited to profile only the individual node. It can't profile cluster of node and Entire Network as a single system.

## 3.9 WOoDSTOCK

WOoDSTOCK is a real-time, on-chip system profiler that Watches Over Data STreaming On Computing element links [15].

*Scope* – WOoDSTOCK profiles system performance by adding monitors to the circuit on the FPGA, which keep track of the communications done between different computing elements (CE) and gauge their utilization. Thus by monitoring communication links between different CEs, it becomes possible to generate a broad picture of system performance, which highlights the computing elements that cause bottlenecks in the design.

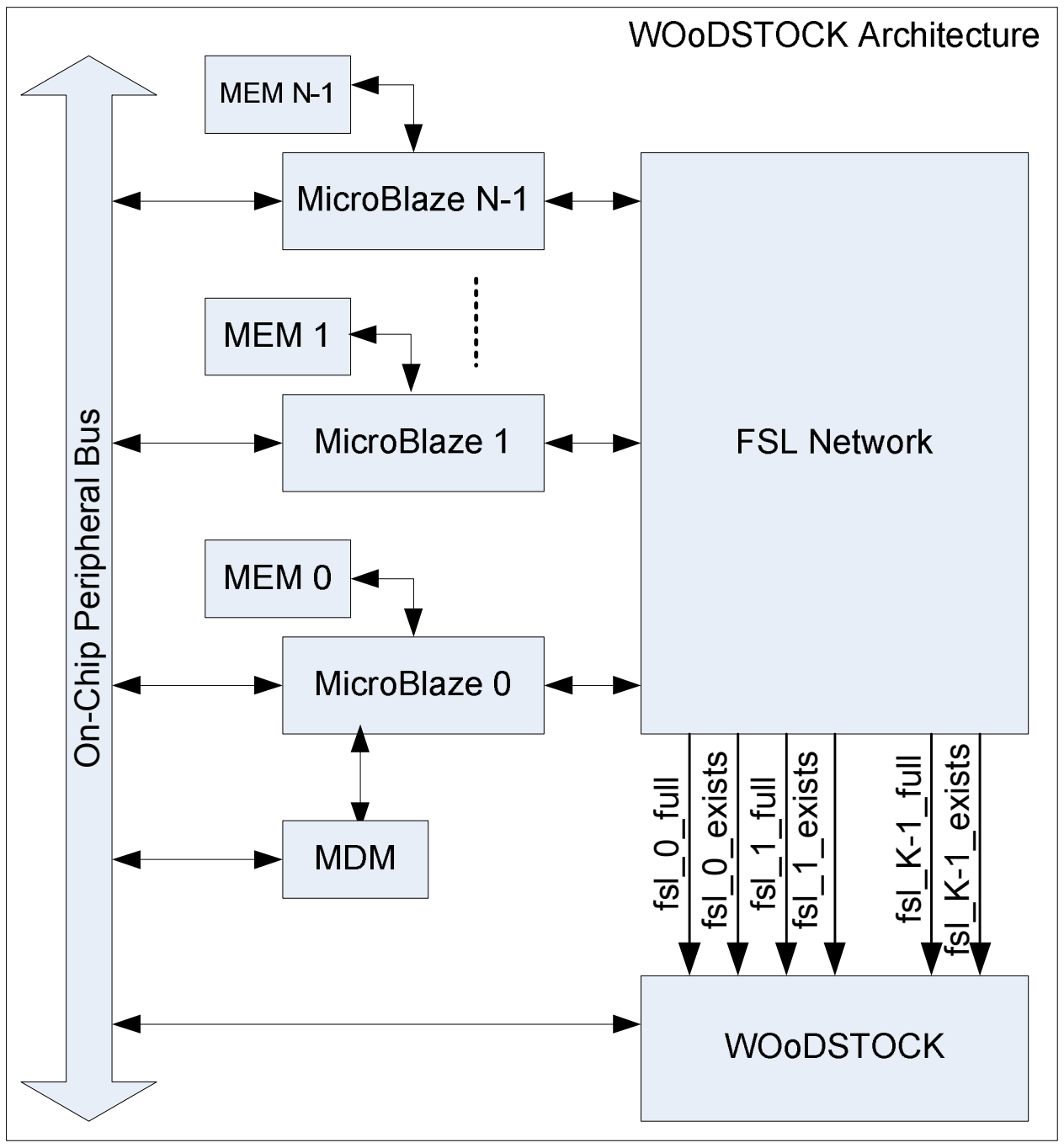

Fig. 11 WOoDSTOCK Architecture

*Architecture* – Connections between WOoDSTOCK and a multi-CE system are shown in Fig.11. A monitor is part of WOoDSTOCK that records the behaviour of traffic on all internal input and output links connected to its CE through internal counters. The content of these counters is used to find the total starving or stalling time for a CE during the profiling period. Processor 0 is the base processor and responsible to determine the run-time for the monitors based on its executing Program Counter (PC_EX). Authors have used Xilinx Virtex2 FPGA and Microblaze soft-core processor to implement and verify the operation of WOoDSTOCK. The performance of the profiler is demonstrated using two applications of different nature: one is based on pipeline and the other based on branching.

*Overhead and Accuracy* – System design is started from an initial configuration which is refined subsequently to a new configuration which is free from stalling and starving. This leads to an average performance improvement by 50% and reduces overall run-time compared to initial configuration. Profiling counters are of 46-bits to allow a maximum system profile period of eight days at 100 MHz. Each counter utilizes 115.8 LUTs and 66.2 Flip-flops for pipeline application and 116 LUTs and 59.8 Flip-flops for branching example.

*Limitation* – WOoDSTOCK does not profile the software application running on the processors just like other profilers discussed in this paper, rather it profiles the communication traffic. Since WOoDSTOCK is unaware about the actual computation being performed on a particular CE, the obvious limitation of WOoDSTOCK is that it could not find the actual cause of bottleneck in designing of CE.

## 3.10 Micro Profiler (µP)

Micro-Profiler is the software application profiler developed to assist designers of Application Specific Instruction-set Processor (ASIP) [12]. There are two instrumentation approaches in use: one is source level and the other is assembly level. The former is better in terms of speed and flexibility where as the latter is good for accuracy. µP is an attempt to fill this gap and offers better accuracy along with reasonable speed. The key idea in µP profiler is to apply source level profiling, yet precisely counting all primitive operations during execution of an application. This become possible by source code level fine grain instrumentation, which inserts extra code in the original application code to collect particular run time statistics without modifying the program semantics.

*Scope* – µP provides designer with important runtime statistics of the application, such as the usage of C operators for different data types, dynamic value ranges of the variables and the constants, coarse performance estimates, etc. for effective pre-architecture exploration to design or customize an ASIP. It can also profile the memory accesses with accuracy close to ISS.

*Architecture* – uP is software based intrusive profiler, which analyzes the code statically to produce the profiling results. In this approach the original C code is transformed to Three Instruction Code. Each line of such code contains at most only one operation. This increases the profiling granularity. LANCE compiler is used to produce such intermediate representation in form of C syntax from original C code. Execution time overhead is minimized by introducing the profiling code at the end of each block only. This code maintains aggregate counters for the entire block, while still being able to provide statement-true information.

Fig.12 shows a piece of C code and corresponding Three Address Code to highlight the limitations of profiling at C source code level. The 3rd statement will be interpreted as a single C statement as like others, where as actually it will be mapped to multiple assembly instructions. Hence C level profiling granularity is too coarse for design of an ASIP ISA. In Three Address Code, all primitive C operations including type cast, pointer scaling, etc. are made more explicit and hence can be profiled like regular operations. All high level memory operations like access of Arrays, Global Variables and Structures are mapped to explicit LOAD/STORE operations via pointers as shown in Fig.12(b).

In spite of code overhead, execution of instrumented code is an order of magnitude faster than that of Instruction Set Simulator. The accuracy in profiling the memory accesses is close to ISS. For an application consisting of n IR statements the cycle count estimate is given by the formula:

$$Cycles = \sum_{i=1}^{n} E(Si) \times W(Si)$$

Where E(Si) and W(Si) are the execution count and weight for statement Si, respectively.

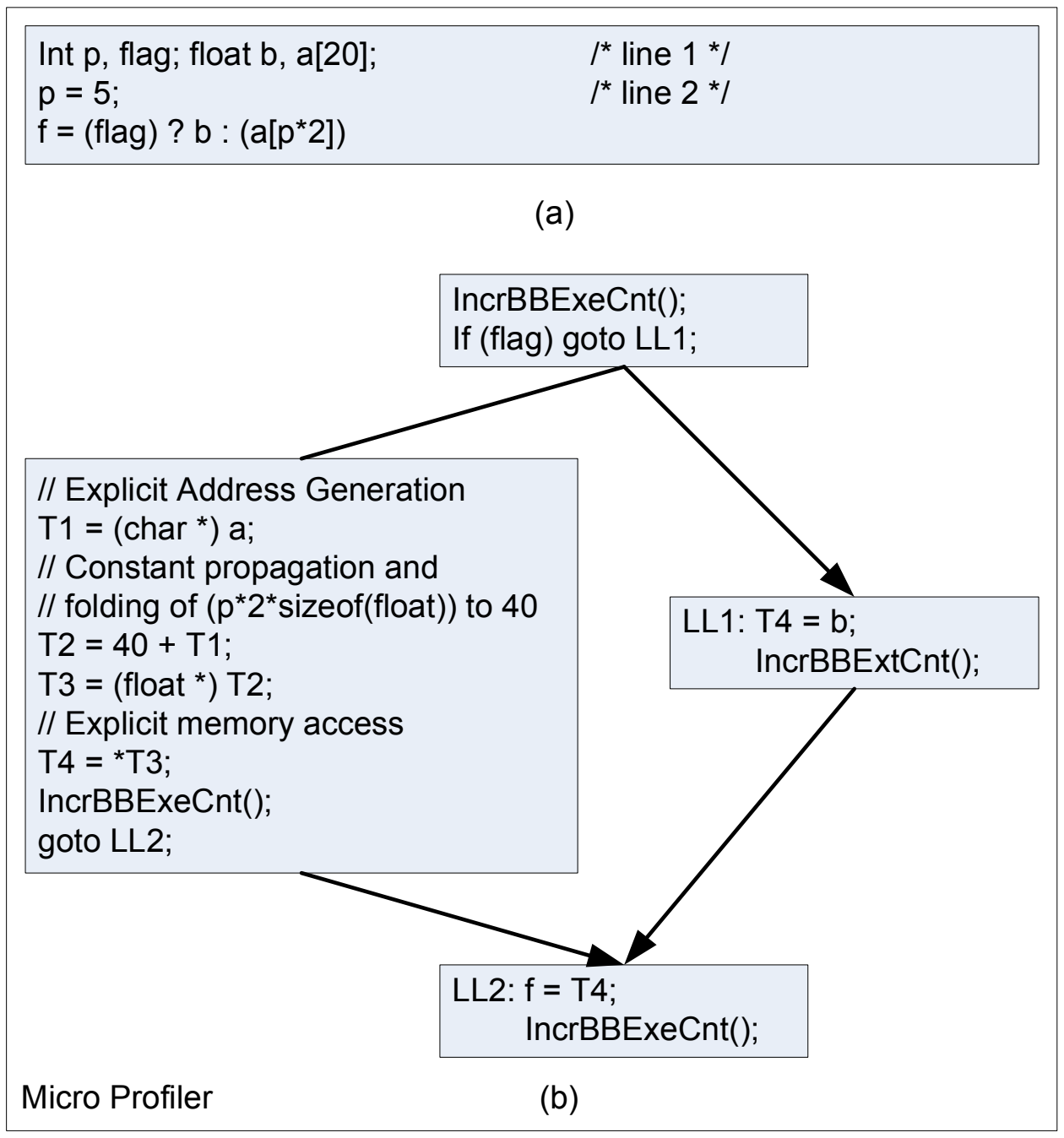

Fig.12 (a) C statements (b) Intermediate representation

*Overhead and Accuracy* – With high level optimization in LANCE compiler, the average deviation reported in cycle count is not more than 11% compared to ISS. Optimization at Intermediate Representation leads to average deviation of 23% in predicting the operator count. For simplicity operators are classified into five basic categories: Arithmetic, Logical, Compare, Load/Store and Multiply/Division/Modulo.

Authors have proposed architecture exploration of ASIP for mpeg3 with the information provided by the uP. Architecture exploration has started with initial reference architecture having 32bit instruction word size, code size 83.08 KB, 18.81K logic gates, no FPU support, clock of 23.60 nS. After repeated cycles of refinement based on profiling information provided by uP, final ISA was proposed with 24 bit instruction word size, code size 11.52 KB, 15.61 K logic gates, with FPU support, clock of 39.68 nS. The original architecture needs 132,649 K cycles/frame. Hence it cannot meet the requirement of playing 38frames/second. However, it has been solved by 410K cycles/frame in finally refined architecture.

Results reflect, slow down of clock by 16.08ns, reduction in code size by 14%, 83% area saving and top of all, final mpeg3 engine is 300 times faster than original architecture.

*Limitation* – The limitation of µP is that it cannot instrument library functions. Hence code comprised of so many library functions may lead to inaccuracy in the profiled data.

## 3.11 Artificial Neural Network Based Techniques

In [13][14], Oyamada et al. has explored and presented Artificial Neural Network as promising higher-level performance estimation technique for embedded software. In higher-level performance estimation, it is not required to fully compile the software application to generate executable binary. Hence there is no execution of the application involved. Profiling information is not captured real-time, this increases the profiling process by many fold. Higher-level prediction is already in use, since a long for simple architectures with no cache, pipeline or branch prediction features. Such performance estimators are modeled using some linear equations. Important advancement in ANN based technique is that it can easily capture the nonlinearity introduced due to features just mentioned. Feed Forward Error Back Propagation neural network is used in this approach.

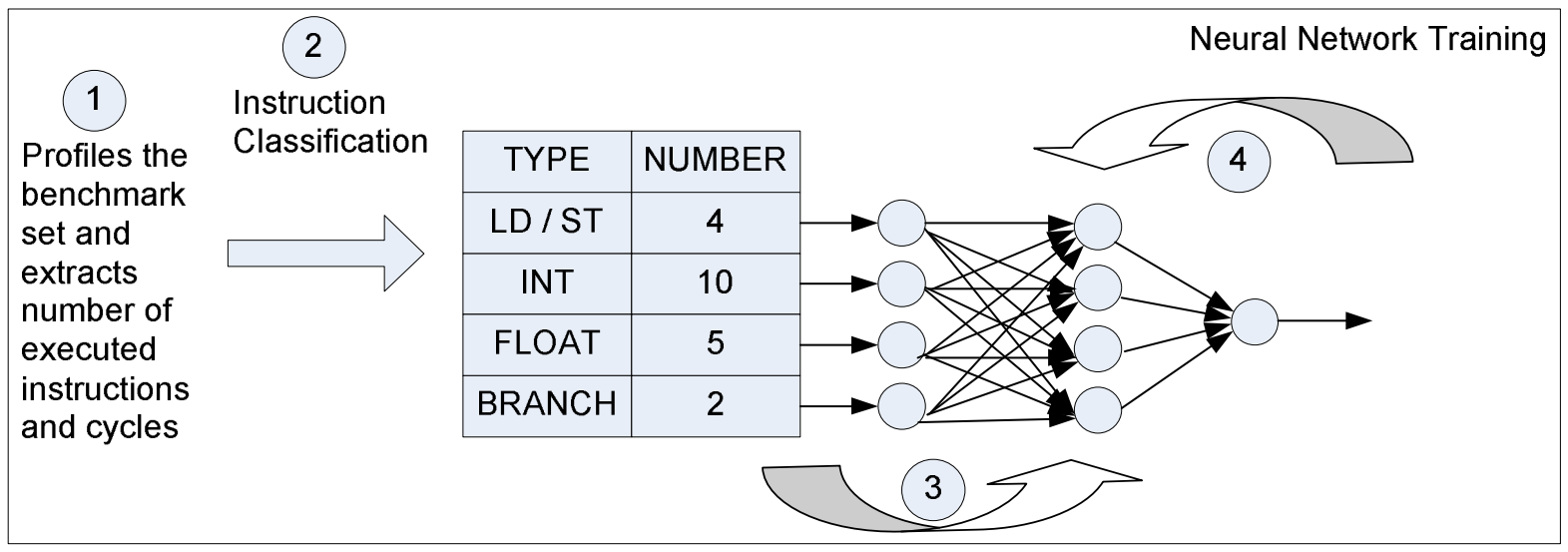

Fig.13 Training Phase of the Estimator

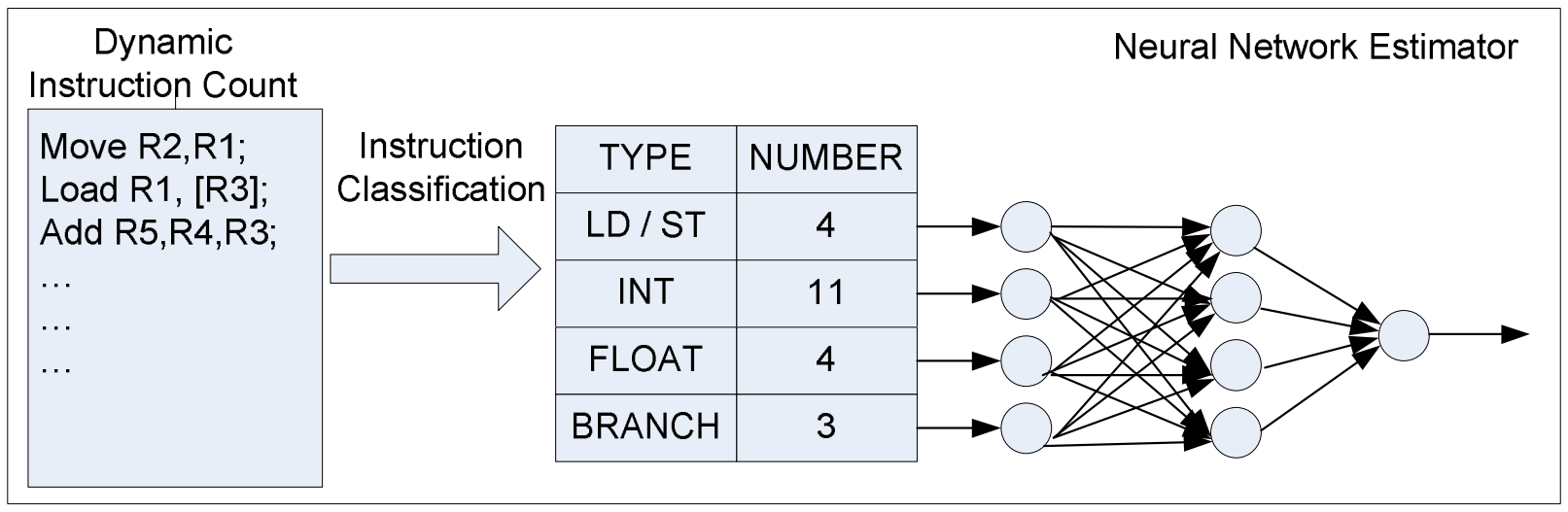

Fig.14 Utilization Phase of the Estimator

*Scope* – The scope of this methodology is found mainly in design space exploration of embedded software, for instance considering various algorithmic alternatives for designing of tasks, assignment of tasks to different computing elements, etc. This is mainly due to its nature to profile the application quickly. This methodology can also be used for automatic classification of the software application, whether it belongs to dataflow dominated or control dominated class.

*Architecture* – This is the software based profiling methodology which is utilized in two steps as shown in Fig.13 and Fig.14: Training and Utilization. The network is trained using the MatLab software on the host machine. Standard benchmark is profiled and number of instructions and cycles executed are identified and classified broadly as shown in Fig.13. This data is then applied as input to train the network. The output of the network is the estimated cycle count which will be compared with the actual profiled number of cycles. The error in cycle count is in turn use to correct the weights of the network. This is how the network is train which is now ready to estimate cycle count for any other application from the similar domain to which training set belongs. Hence a library of trained neural networks can be prepared for different processor architectures and applications of different domain for ready use in future for fast design space exploration.

After the training phase, the estimation tool is ready to be used. Fig.14 presents the main steps in this phase. An application is compiled for a given target processor, and the number of executed instructions of each type is obtained by a dynamic instruction count. These counts are presented to the neural network, so that it can estimate the total number of cycles consumed by the application.

*Overhead and Accuracy* – The approach has been validated using PowerPC 750, ADSP21xx and a Java Microcontroller using a set of 39 benchmarks. Authors have demonstrated use of this technique using PowerPC 750 as target architecture and speed up to 190 times is reported compared to cycle accurate simulator with reasonably acceptable mean error of 7.90%. The mean error is further reduced to 6.48% by using domain specific ANN to estimate the performance. This is equivalent to execution cycle accuracy of 92.10% and 93.52%, respectively.

*Limitation* – However the approach is fast to estimate the cycle count once the network is trained, the training phase is considerably long for every new neural network. Hence it is required to identify some methods that can expedite the process of training the neural network. Every application to be profiled, which is written in higher level language needs to be compiled to count instructions of different class.

# 4. Discussion

A summary of all eleven presented embedded software profiling methodologies based on the classification criteria introduced in Section II is given in Table II.

As can be seen, most of the methodologies share many common characteristics. For example, most are Nonintrusive, Low-level and dynamic in nature. Most do not support profiling of multitask applications. If we talk about accuracy, SnoopP profiler gives nearly 100% accurate profiling information. It is so because it pickup program counter value without any software overhead. However it selectively profiles application segments and needs resynthesis of profiler for every new configuration to be profiled. Whereas gprof incurs maximum error and it seems to be more appropriate for profiling general computing software rather than embedded software.

TABLE II
CLASSIFICATION AND COMPARISION OF DIFFERENT PROFILING APPROACHES

| Sr. No. | Approach | Intrusive | mplementatio | | Abstraction | | Analysis | | Task Support | | Accuracy<br>ECA – Execution Cycle Accurate<br>CCA – Clock Cycle Accurate<br>CAS – Cycle Accurate Simulator<br>LCA – Loop Count Accurate | Overhead<br>PO – Performance Overhead<br>EO – Execution Overhead<br>AO – Area Overhead<br>LGO – Logic Gate Overhead |
|---|---|---|---|---|---|---|---|---|---|---|---|---|
| | | | SW Based | FPGA Based | Low Level | High Level | Static | Dynamic | Single Task | Multi Task | | |
| 1 | GNU's *gprof* | • | • | | | | | • | • | | Not Mentioned | 20.37% PO over Address Tracer [5] |
| 2 | SnoopP | | | • | • | | | • | • | | 99.94% ECA | Large Area and 0.01% EO |
| 3 | Airwolf | ○ | | ○ | • | | | • | • | | 66.20 % ECA over gprof | 0.23% PO over Address Tracer [5] and 41% reduction in PO over gprof |
| 4 | Address Tracer | | | • | • | | | • | • | | Close to Airwolf | Large Area |
| 5 | FLAT | | SoC based synthesized using 0.18 uN USMC Library | | • | | | • | • | | 80% ECA | 2.4% power overhead<br>6.68 to 12.8 % AO |
| 6 | DAProf | | | | • | | | • | • | | 98% Loop Count Accurate and 95% Execution Count Accurate | 20% AO compared to ARM9 |
| 7 | MPPA | ○ | | ○ | • | | | • | | • | 93.41% ECA | 0.66% LGO |
| 8 | DPOP | • | • | | • | | | • | | • | 83 % improvement in design cost over hand optimized sensor node | 1.7 to 2.4 % Code Overhead<br>7.9 to 14.8 % Traffic Overhead<br>0.01 to 0.06 mAH Energy Overhead |
| 9 | WOoD STOC K | | | ● | ● | | | ● | | ● | 50% Performance Improvement | Average 116 LUTs and 60 Flip-flops per counter of profiler |
| 10 | µP | ● | ● | | | ● | ● | | - NA - | | 300 Times Performance Gain | Not Reported |
| 11 | ANN Based | | ● | | | ● | ● | | - NA - | | 92.10 % to 93.52% CCA<br>190 times speed up over CAS | Not Mentioned |

Most of the approaches are profiling the software application running on the processor core with two exceptions; DPOP and WOoDSTOCK. DPOP framework is mainly meant for optimizing the traffic/energy/code size in every node of wireless sensor network. WOoDSTOCK is the approach in which performance of the system is gauge by monitoring the communication traffic between computing elements of the system. Similarly there is one more dimension in which significant research has been done, which is profiler based on monitoring the frequently executing loops.

Vahid et al. had given this concept in FLAT profiler which has been further extended by Nair et al. in developing DAProf. The most recent approach is DAProf, which greatly enhances the performance of WARP processors. Warp processing is the technology that dynamically detects the potential kernels of an executing binary which are suitable for HW implementation without any designer efforts and hence addresses the critical issue of HW/SW partitioning. Such dynamic profilers are very suitable, especially for dynamic optimization approaches.

Micro Profiler is another exception which is specially meant for architecture optimization of ASIP. Performance of ASIP can be greatly improved with optimized hardware by profiling every new refined architecture with uP successively.

Another important point to be focused is that most of the embedded systems today are realized on a single FPGA in SoC form. Hence FPGA based profilers seem to be more efficient for profiling application running on a soft or hard core processor synthesized on an FPGA. However such profilers are quite processor architecture dependent and not fast enough for rapid system level design space exploration. Hence people are motivated towards Artificial Neural Network based profilers. Such profilers are very fast and accuracy offered is also reasonably good. Such profilers are quite helpful for fast design space exploration at system level.

# 5. Conclusion

The survey presented in this paper may be greatly helpful to the researchers and engineers who want to identify a most suitable embedded software profiler based on their system or application requirement. The survey also gives numerous new research directions which are represented as limitations in discussion of each approach. In this survey an attempt is made to cover all possible approaches addressed by different authors in the domain of embedded software profiling in last seven to eight years. An exhaustive comparison based on various criterions is also provided for ready reference.

A special attention is expected to develop profiler for OS based or multitasking embedded software.

Another area that needs to be addressed is to compile a library of trained neural networks for different processor architectures. Such neural networks can be trained by different domain specific applications to increase their accuracy. This kind of library based approach can increase the high level design space exploration and processor choice by many folds.

## Authors

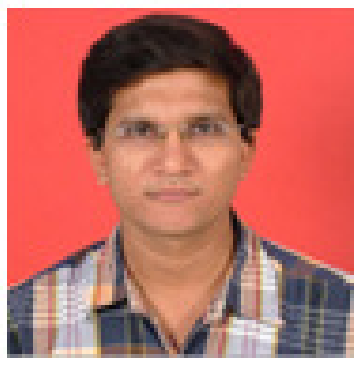

**Rajendra Patel** received his Bachelor of Engineering (Electronics and Communication Engineering) degree in 2001 from the University of Bhavnagar, Bhavnagar, Gujarat, India. He received degree of Master of Technology (VLSI and Embedded **System** Design) in 2009 from National Institute of Technology, Bhopal, Madhya Pradesh, India. He is currently pursuing his Ph.D. from the same institute. He had been accorded with distinction in both of his degrees. His research interests include Embedded Software Development, Hardware-Software Co-Design, Reconfigurable Systems and Digital System Design.

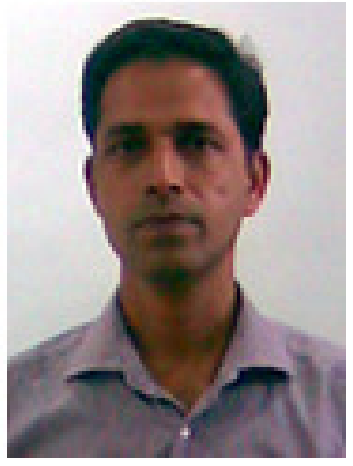

**Arvind Rajawat** has received his BE in Electronics & Communication Engineering from Government Engineering College, Ujjain, India and ME in Computer Engineering from SGSITS Indore, India. He earned his PhD in the area of Communication Architecture Exploration in Codesign from MANIT Bhopal, India. He is having teaching experience of more than 20 years and he is currently working as Associate Professor in Electronics and Communication Engineering, MANIT, Bhopal. His main area of interest and research are Real Time Embedded System Design, Hardware Software Codesign, Architecture Exploration, etc.